\documentclass[11pt]{article}

\usepackage{latexsym}
\usepackage{amsfonts}
\usepackage{amssymb}
\usepackage{amsmath}

\font\tenmsbm=msbm10 scaled 1200
\font\sevenmsbm=msbm9
\newfam\msbmfam
\textfont\msbmfam=\tenmsbm \scriptfont\msbmfam=\sevenmsbm

\makeatletter
\@addtoreset{equation}{section}
\makeatother

\newcommand{\eref}[1]{(\ref{#1})}

\def\be{\begin{equation}}
\def\ee{\end{equation}}
\def\ba{\begin{align}}
\def\bet{\begin{tabular}}
\def\eet{\end{tabular}}
\def\pa{\partial}

\def\a{\alpha}
\def\bt{\beta}
\def\g{\gamma}
\def\G{\Gamma}

\def\dl{\delta}
\def\la{\lambda}

\def\m{\mu}
\def\n{\nu}
\def\ra{\rightarrow}
\def\vp{\varphi}
\def\ul{\underline}

\def\vp{\varphi}
\def\fr{\frac}

\long\def\symbolfootnote[#1]#2{\begingroup%
\def\thefootnote{\fnsymbol{footnote}}\footnote[#1]{#2}\endgroup}

\begin{document}

\begin{titlepage}

\begin{flushright}
Preprint DFPD/2012/TH25\\
December 2012\\
\end{flushright}

\vspace{1truecm}

\begin{center}

{\Large \bf The Lienard-Wiechert field of accelerated massless charges} \vskip0.3truecm

\vspace{1.5cm}

Francesco Azzurli$^1$\symbolfootnote[1]{francesco.azzurli@studenti.unipd.it}
and Kurt Lechner$^2$\symbolfootnote[2]{kurt.lechner@pd.infn.it}

\vspace{1.5cm}
$^1${\it Scuola Galileiana di Studi Superiori - Universit\`a degli Studi di Padova, Italy}

\vspace{1cm}
$^2${\it Dipartimento di Fisica e Astronomia, Universit\`a degli Studi di Padova, Italy

\smallskip
and

\smallskip
INFN, Sezione di Padova,

Via F. Marzolo, 8, 35131 Padova, Italy}

\vspace{1.5cm}

\begin{abstract}

We determine for the first time the electromagnetic field generated by a generic massless accelerated charge, solving exactly Maxwell's equations. This result may shed new light on the possible existence of such particles in nature.

\vspace{0.5cm}

\end{abstract}

\end{center}
\end{titlepage}

\newpage

\baselineskip 5 mm


\section{Introduction}

Massless charged particles do not seem to exist in nature, the lightest known charged particle being the electron. Apparently there is no fundamental theoretical principle that forbids their existence, although the consistency of relativistic quantum field theories describing massless charges and the related cancellation of infrared divergences are still open problems \cite{LM}. Surprisingly the consistency of those particles at the classical level is seemingly an open question and very little is known about their physical features: the fact that they travel with the velocity of light introduces mathematical, as well as physical, singularities that make the most standard basic instruments of Functional Analysis to solve Maxwell's equations -- like the Green function method -- difficult to apply. Basically what is known about classical massless charges is merely the electromagnetic field generated by a (free) charge in {\it uniform linear} motion, that is given by  a $\dl$-like shock-wave supported on a plane orthogonal to the trajectory, travelling at the speed of light \cite{RR,AE}:
\be\label{free}
F^{\m\n}_0(x)=\fr e {2\pi} \fr{u^\m x^\n-u^\n x^\m}{x^2}\,\dl(ux).
\ee
Here  $e$ is the charge of the particle, $u^\m$ its four-velocity subject to the lightlike condition $u^2=0$, and $(ux)$ stands for $u^\m x_\m$.

In this paper we solve for the first time Maxwell's equations for a particle moving with the velocity of light along a generic accelerated trajectory and determine explicitly the electromagnetic field it produces. For the sake of simplicity we consider only {\it bounded} trajectories, but the new method we develop applies equally well to unbounded ones. Since the current density of a point-like particle  is a Dirac  $\dl$-function, the solution makes sense only in the space of distributions. The corresponding electromagnetic field ${\cal F}^{\m\n}$, for which we give an analytical expression in \eref{emf}, is Lorentz-covariant and respects causality and represents thus the generalization of the celebrated Lienard-Wiechert field from timelike to lightlike worldlines. The relevance of this result is first of all of conceptual character, since it proves that Maxwell's equations admit (exact) solutions in the space of distributions even if the worldlines are lightlike, although conventional methods fail in this case, as observed above.
The knowledge of the exact field produced by an accelerated massless charged particle should eventually allow to settle the question whether or not a consistent classical radiation theory for such particles -- preserving total energy and momentum -- can be formulated; for preliminary analyses see e.g. \cite{XY}-\cite{Y}. The knowledge of this field represents furthermore the basis for a systematic analysis of possible {\it interactions} of this kind of particles, while the field \eref{free} describes only free particles. The results presented in this paper could thus shed new light on the open question of the possible existence of massless charged particles in nature, even though a definitive answer to this question can arise only in the framework of quantum field theory.
Eventually the method developed in this paper to derive exact solutions for linear partial differential equations in the space of distributions can be applied also to other physically relevant problems, like the field produced by a massless dyon or the {\it linearized} gravitational field produced by a massless particle, like the photon. Detailed proofs and further developments will be presented elsewhere \cite{AL}.

\section{Heuristic analysis and the singularity surface $\G^\m$}

A massless charged particle travels with the speed of light and,
as long as its energy and mass are not directly involved in the physical process one considers, it appears natural to regard it as some limiting case of a particle that travels with speed $V<1$.  With this respect a speed $V<1$ in some sense plays a role complementary to a mass $m>0$ used frequently in quantum field theory to regularize infrared divergences.

{\it Timelike worldlines.} Our starting point is therefore the standard Lienard-Wiechert field for a timelike worldline, that we write as
\be\label{ftot}
F^{\m\n}=C^{\m\n}+R^{\m\n},
\ee
where the Coulomb and radiation fields are given respectively by
\ba
\label{vmn}
C^{\m\n}(x)&=\fr{e\,U^2}{4\pi (UL)^3}\left(L^\m U^\n-L^\n U^\m\right),\\
R^{\m\n}(x)&=\fr{e}{4\pi (UL)^3}\,L^\m\left[(UL)W^\n-(WL)U^\n\right]-(\m\leftrightarrow\n).
\label{amn}
\end{align}
Here we parametrized the particle's worldline $Y^\m\equiv Y^\m(\la)$ with an arbitrary parameter $\la$ and defined $L^\m(x)=x-Y^\m$, $U^\m=dY^\m/d\la$, $W^\m=dU^\m/d\la$. We adopt the convention $(UL)=U^\m L_\m$ and similarly for $(WL)$. In  formulae \eref{vmn} and \eref{amn}  $\la$ is evaluated at its ``retarded'' value $\la(x)$, uniquely determined by the conditions
\be\label{ret}
L^2=\left(x-Y(\la)\right)^2=0, \quad L^0=x^0-Y^0(\la)>0.
\ee
Notice that the square of the ``four-velocity'' $U^\m$ is given by
\be\label{u2}
U^2=\left(1-V^2\right)\left(\fr{dt}{d\la}\right)^2,
\ee
where $V\equiv| \vec V|=| d\vec Y/dt|$ is the speed of the particle. The above fields satisfy the Maxwell equations
\be\label{maxtl}
\pa_{[\m} {F}_{\n\rho]}=0,\quad\quad \pa_\m F^{\m\n}=J^\n\equiv e\int\dl^4(x-Y) \,dY^\n
\ee
in the {\it sense of distributions}\,: due to the presence of  $\dl$-like currents this is the only framework where these equations make sense.
In particular the fields $C^{\m\n}$ and $R^{\m\n}$ satisfy separately the Bianchi identities
\be\label{maxbian}
\pa_{[\m} R_{\n\rho]}=0=\pa_{[\m} C_{\n\rho]}.
\ee

{\it Lightlike worldlines.} We consider now an arbitrary  spatially  {\it bounded} lightlike worldline $y^\m\equiv y^\m(\la)$ and search for the causal solution of the associated Maxwell equations
\be\label{maxll}
\pa_{[\m} {\cal F}_{\n\rho]}=0,\quad\quad \pa_\m {\cal F}^{\m\n}=j^\n\equiv e\int\dl^4(x-y)\, dy^\n.
\ee
In analogy to the timelike worldline $Y^\m$ we define $u^\m=dy^\m/d\la$ and $w^\m=du^\m/d\la$ where, since the wordline is lightlike, $u^2=0$. For future reference we introduce also a retarded parameter $\la^*(x)$ defined analogously to \eref{ret}:
\be\label{lstar}
l^\m(x)\equiv x-y^\m(\la^*),\quad l^2=0,\quad l^0>0.
\ee
Loosely speaking we will derive the solution of equations \eref{maxll} considering a conveniently deformed timelike worldline $Y^\m(\la)$ that in an appropriate limit -- to be specified below --  approaches the given lightlike worldline $y^\m(\la)$. In this limit we have $\la(x)\ra\la^*(x)$,  $U^2\ra u^2=0$, $(UL)\ra (ul)$ and $V(t)\ra v(t) =1$, where $\vec v(t)=d\vec y/dt$ is the velocity of the lightlike worldline. For what concerns the fields \eref{vmn} and \eref{amn}, since in this limit $U^2\ra 0$ the field
$C^{\m\n}$ appears to vanish, while $R^{\m\n}$ seems to admit a finite limit as it does not involve any power of $U^2$. Both these conclusions are, however, of purely conjectural character due to the presence of zeros in the denominators of \eref{vmn} and \eref{amn}. These denominators are, in fact, powers of the scalar product $(UL)$, that tends to $(ul)$. But while $(UL)$, given \eref{ret}, vanishes only on the particle's worldline $x=Y^\m(\la)$, i.e. at fixed time $t$ only at the particle's position, the scalar product $(ul)$ vanishes along the {\it physical singularity surface} $x^\m=\G^\m(b,\la)$  defined by\footnote{For a uniform motion -- which is necessarily {\it unbounded} -- this surface collapses to the worldline of the particle and the following argument does not apply. In this case the Coulomb field admits actually the non-zero limit \eref{free}. For a qualitative description of the singularity curve \eref{3sing} in a cyclotron motion see \cite{Y}.}
\be\label{gg}
\G^\m(b,\la)\equiv y^\m(\la)+b\,u^\m(\la), \quad b>0.
\ee
At fixed time $t$ the spatial component of this surface is a Dirac-like string attached to the particle given by
\be\label{3sing}
\vec \gamma(b,t)=\vec y(t-b)+b\,\vec v(t-b),\quad b>0.
\ee
The fact that $(ul)$ vanishes on $\G^\m$ follows from the relations $l^2=(x-y(\la))^2=0$, $x^0-y^0(\la)>0$ and $u^2(\la)=0$. On the singularity surface $\G^\m$, that will play a crucial role in what follows, the denominators of \eref{vmn} and \eref{amn} tend therefore to {\it zero}. This means that, {\it if the above envisaged limits of $C^{\m\n}$ and $R^{\m\n}$ in some sense exist}, these limits could involve a ``special'' tensor {\it supported on $\G^\m$}.  Due to Lorentz-covariance, reparametrization invariance ($\lambda\ra \lambda'$, $b\ra (d\la'/d\la)\,b)$, antisymmetry in $\m$ and $\n$ and for dimensional reasons, this tensor can only be proportional to
\be\label{CG}
P^{\m\n}(x)\equiv e\int_{-\infty}^\infty d\la\int_0^\infty b  \left(u^\m w^\n-u^\n w^\m\right)\dl^4(x-\G(b,\la))\,db.
\ee
As a simple calculation shows this field entails furthermore another interesting property, that is
\be\label{dcmn}
\pa_\m P^{\m\n}=j^\n,
\ee
where $j^\n$ is the lightlike current defined in \eref{maxll}. For $V\ra 1$ the limit of the Coulomb field $C^{\m\n}$ could thus only be proportional to $P^{\m\n}$ since, as observed above, in the complement of $\Gamma^\m$ it goes to zero. However, $C^{\m\n}$ satisfies the Bianchi identity $\pa_{[\m} C_{\n\rho]}=0$  and so also its (possible) limit $k P^{\m\n}$ must satisfy it. But since, by inspection, $\pa_{[\m} P_{\n\rho]}\neq 0$ we conclude that $k=0$: this implies that for a lightlike bounded worldline the {\it Coulomb field must vanish identically}. This argument does clearly not apply to the limit of the radiation field $R^{\m\n}$, that could -- and will indeed -- develop a contribution proportional to $P^{\m\n}$.

\section{A rigorous strategy}

To establish a well defined limiting procedure we have first of all to establish a feasible timelike {\it regularized} wordline $Y^\m(\la)$ for a given arbitrary lightlike bounded worldline $y^\m(\la)$. As will be clear in a moment, we need this regularized worldline to depend on a {\it constant} regularization parameter $V$.  As simple as it may look, we define the regularized timelike wordline by
\be\label{reg}
Y^0(\la)\equiv\fr { y^0(\la)}V,\quad \quad \vec Y(\la)\equiv\vec y(\la), \quad 0<V<1.
\ee
The regularized {\it orbit} is thus the same as the orbit of $y^\m(\la)$, but the velocity of the particle is now $\vec V(t)\equiv d\vec Y/d Y^0= V\vec v(t)$ and hence its speed is a {\it constant} less than unity: $|\vec V(t)|=V<1$. The regularized  wordline is thus timelike and the regularized field $F^{\m\n}$ is given by \eref{vmn} and \eref{amn} and depends on the constant parameter $V$. As this field is a distribution, it makes sense to analyse its limit as $V$ tends to 1 {\it in the sense of distributions.}

{\it Distributional limit.} We want now to construct a solution of  equations \eref{maxll} in the distributional sense. In what follows we indicate the {\it pointwise} limit of a function with the standard symbol $\lim_{\,V\ra 1}$, while we write the {\it limit in the sense of (tempered) distributions} as ${\rm Lim}_{\,V\ra 1}$. First of all we notice the trivial distributional limit, see equations \eref{maxtl} and \eref{maxll},
\be\label{jlim}
{\rm Lim}_{\,V\ra 1}J^\m =j^\m.
\ee
Since in the space of distributions derivatives are continuous operations, partial derivatives commute always with the distributional limit ${\rm Lim}_{\,V\ra 1}$. Therefore -- and this is one of the key points of this paper -- \ul{\it if} the distributional limit of the field \eref{ftot}
\be\label{limf}
{\rm Lim}_{\,V\ra 1}\,F^{\m\n}\equiv {\cal F}^{\m\n}
\ee
exists, applying the operation ${\rm Lim}_{\,V\ra 1}$ to all equations in \eref{maxtl}
and interchanging the limits with the derivatives, one can conclude that the field {\it ${\cal F}^{\m\n}$ satisfies automatically the Maxwell equations \eref{maxll}}. In the rest of the paper we outline how the limit \eref{limf} can be evaluated explicitly.

We conclude this section noting that, thanks to \eref{u2}, in the complement of $\G^\m$ the pointwise limit of \eref{ftot} amounts to the function
\be\label{camn}
\lim_{\,V\ra 1}F^{\m\n}(x)=\lim_{\,V\ra1}R^{\m\n}(x)=\fr{e}{4\pi (ul)^3}\,l^\m[(ul)w^\n-(wl)u^\n]-(\m\leftrightarrow\n)\equiv {\cal R}^{\m\n}(x).
\ee

\section{The limiting procedure}

{\it Limit of the Coulomb field.} We discuss first the distributional limit of the Coulomb field ${\rm Lim}_{\,V\ra1}\,C^{\m\n}$ in \eref{vmn}. The evaluation of this limit requires to evaluate the ordinary limits $\lim_{\,V\ra 1} C^{\m\n}\,(\vp)= \lim_{\,V\ra 1}\int C^{\m\n}(x)\,\vp(x)\,d^4x$, where $\vp(x)$ is an arbitrary test function  belonging to the Schwartz-space ${\cal S}(\mathbb{ R}^4$). For the reasons reported above these limits are conjectured to be all zero: using the techniques that we will develop below for the radiation field, an explicit calculation \cite{AL} shows indeed that these limits vanish\footnote{In this proof it is crucial that the trajectory is {\it bounded}. We stress that for a linear uniform trajectory the limit \eref{limv} gives indeed the non vanishing result \eref{free}.}. We have thus
\be\label{limv}
{\rm Lim}_{\,V\ra1}\,C^{\m\n}=0.
\ee
{\it Limit of the radiation field.} In the remainder of this section we establish the existence of the distributional limit ${\rm Lim}_{\,V\ra1}R^{\m\n}$ and evaluate it explicitly, the result being \eref{emf}. We perform the analysis explicitly for the electric component $R^{i0}$ of the field; its magnetic component $R^{ij}$ can be analyzed in an identical fashion. We must thus apply this field to a test function and consider the quantity $R^{i0}(\vp)=\int R^{i0}(x)\,\vp(x)\,d^4x$. To analyze this integral it is first of all convenient to disentangle  the retarded parameter $\la(x)$ -- transforming it in a ``free'' parameter -- inserting the $\dl$-function identity
$$
2\int(UL)H(L^0)\,\dl\big( L^2\big)\,d\la=\int\dl(\la-\la(x))\,d\la=1,\quad L^\m\equiv x^\m-Y^\m(\la),
$$
where $H(\,\cdot\,)$ stands for the Heaviside  function.
After that we perform the shift of variables $x^\m\ra x^\m+Y^\m(\la)$, integrate the $\dl$-function over $x^0$ and use then reparametrization invariance to choose as time variable $t\equiv Y^0(\la)$. Eventually we insert the regularized worldline \eref{reg} and perform the rescaling $t\ra t/V$. In this way from \eref{amn} we get
\be\label{ai0}
R^{i0}(\vp)=\fr {eV} {4\pi}\int K^i(x)\,\vp\left(\tfrac t V +r,\vec x+\vec y\right)dtd^3x,\quad r=|\vec x|,
\ee
where
\be\label{fi}
K^i(x)\equiv
\fr
{(\vec a\cdot\vec x)\left(x^i-rVv^i\right)-r(r-V(\vec v\cdot\vec x))\,a^i}
{r\left(r-V(\vec v\cdot\vec x)\right)^2}.
\ee
In expressions \eref{ai0} and \eref{fi} the kinematical variables $\vec y$, $\vec v$ and $\vec a$ are respectively the position, velocity and acceleration of the lightlike worldline and are {\it evaluated at the time $t$}. We must now evaluate the limit of \eref{ai0} as $V\ra 1$. As long as $V<1$ the denominator of \eref{fi} vanishes only in $r=0$, but for $V=1$ (at fixed $t$) it vanishes along the half-line $\vec x(b)=b\, \vec v $, $b>0$, leading to a non integrable singularity in \eref{ai0}.
This half-line is precisely the image of the singularity curve $\vec \g$ in \eref{3sing}. To isolate  this line we perform in \eref{ai0} a final change of the space-coordinates   $\vec x\leftrightarrow (b, q_1,q_2)$, introducing at fixed $t$ a basis $\vec N_
\a$ orthogonal to $\vec v\equiv \vec v(t)$:
\be\label{xbq}
\vec x=b\,\vec v+q_\a\vec N_\a, \quad \vec N_\a\cdot\vec N_\bt=\dl_{\a\bt},\quad \vec N_\a\cdot\vec v=0,\quad \a,\bt=1,2.
\ee
In these coordinates the singularity curve $\vec \g$ is represented  simply by the conditions $q_\a=0$ and $b>0$. Taking into account that $v=1$,  $\vec a\cdot\vec v=0$ and that the Jacobian of the transformation \eref{xbq} is 1, expression \eref{ai0} can be rewritten as
\be\label{ai01}
R^{i0}(\vp)=\fr {eV} {4\pi}\int\left (G^i+H^i\right)\vp\left(\tfrac t V +r,b\,\vec v+q_
\a\vec N_\a+\vec y\right)dt\,db\,d^2q,
\ee
where $r=\sqrt{b^2+q^2}$,  $q^2=q_1^2+q_2^2$ and we introduced the functions
\ba\label{gi}
G^i&=-\fr{(1-V^2)(r+Vb)b^2a^i}
{\left(q^2+(1-V^2)b^2\right)^2},\\
H^i&=\fr{\Pi_{\a\bt}N_\a^iN_\bt^j\,a^j}
{r(r-Vb)^2}
+\frac{q_\a(\vec a\cdot\vec N_\a)(b-Vr)v^i}{r(r-Vb)^2}
- \fr{q^2a^i}{2r(q^2+(1-V^2)b^2)},
\label{hi}
\end{align}
where
$$
 \Pi_{\a\bt}\equiv q_\a q_\bt-\fr{q^2}{2}\,\dl_{\a\bt}.
$$
Under the limit $V\ra 1$ both integrals in \eref{ai01} converge now separately. The delicate point is represented by the ({\it a priori} non integrable) divergences arising at $q=0$ when $V$ goes to $1$: the interchange of the limit $V\ra 1$ with the integral over $d^2q$ must therefore be handled with care. In the $G^i$-integral one can first perform the rescaling $q_\a\ra\sqrt{1-V^2}\, q_\a$ and resort then to the {\it dominated convergence theorem} to move the limit $V\ra 1$ inside the $q$-integral. The resulting integral over $d^2q$ is then convergent, and elementary, and the test function $\vp$ gets eventually evaluated at $q=0$. For $b<0$ this integral is actually zero, because in this case under the above rescaling of $q_\a$ the term $(r+Vb)$ vanishes as $V\ra 1$.

For what concerns instead the $H^i$-integral, in the second and third terms of \eref{hi} the limit can be taken trivially under the $q$-integral sign. On the contrary the first term of \eref{hi} is  characterized by the singular behavior of its denominator, that for $b>0$ under $V\ra 1$ for small $q$ behaves as $1/(r-Vb)^2\ra  1/(r-b)^2\sim 1/(q^2)^2$, while its numerator is quadratic in $q$. For $V\ra 1$ the corresponding integral seems thus logarithmically divergent. However, if before taking the limit $V\ra 1$ one performs {\it first} the integral over the polar angle $\vartheta$ of the $q$-plane and {\it afterwards} the integral over its modulus $q=\sqrt{q_1^2+q_2^2}$, the logarithmic divergence cancels because by symmetric integration the integral over $\vartheta$ of the matrix $\Pi_{\a\bt}$ vanishes. In the so rearranged $ H^i$-integral -- enforcing again the dominated convergence theorem -- one can take the limit $V\ra 1$ under the $q$-integral sign and the resulting integral is {\it conditionally convergent}\footnote{In the most simplest case an integral is said to be {\it conditionally convergent}, if it converges once one specifies a specific order of the integrations over the variables.} in the sense just specified. In this way the limit of \eref{ai01}, that thanks to \eref{limv} corresponds to the $i0$-component of the distributional limit \eref{limf}, becomes eventually the sum of two terms:
\ba\label{ai02}
&{\cal F}^{i0}(\vp)=
\lim_{V\ra1}R^{i0}(\vp)=-\fr e2\int_{-\infty}^\infty dt\int_0^\infty a^i\,\vp\left(t+b,b\,\vec v+\vec y\right)b\,db\\
&+\fr {e} {4\pi}\int\left(\fr{\Pi_{\a\bt}N_\a^iN_\bt^j\,a^j}
{r(r-b)^2}
-\frac{q_\a(\vec a\cdot\vec N_\a)v^i}{r(r-b)}
-\fr{a^i}{2r}\right)
\vp \left(t+r,b\,\vec v+q_\a\vec N_\a+\vec y\right)dt\,db\,d^2q,
\label{ai03}
\end{align}
where it is understood -- we repeat -- that the $q$-integral in \eref{ai03} is conditionally convergent. The term in \eref{ai02} is supported on the singularity line \eref{3sing} and by inspection one sees that it amounts precisely to $\tfrac12\, P^{i0}(\vp)$, see \eref{CG}. To the distribution in \eref{ai03} one can {\it formally} apply all the coordinate transformations performed so far ``backwards'', and in doing so one discovers that in the complement of $\G^\m$ it can be written as $\int{\cal R}^{i0}(x)\,\vp(x)\,d^4x$, where the function ${\cal R}^{i0}(x)$ is defined in \eref{camn}. This result is obviously not surprising, since in regions not containing singularities distributional and pointwise limits are equivalent. Repeating the same analysis for the magnetic components ${R}^{ij}$ of \eref{amn}, and taking into account that the above analysis holds for an arbitrary test function $\vp$, one concludes that the limit \eref{limf} exists and can be written as
\be\label{emf}
{\cal F}^{\m\n}=\fr 1 2\,P^{\m\n}+{\cal P}({\cal R}^{\m\n}).
\ee
The ``principal part'' symbol ${\cal P}({\cal R}^{\m\n})$ indicates the distribution defined in terms of the function ${\cal R}^{\m\n}(x)$  given in \eref{camn}, through the $q$-integration procedure around the singularity curve $\vec \gamma$  prescribed above.

\section{Analysis and comments}

We conclude the paper with an analysis of the most peculiar and salient features of the expression ${\cal F}^{\m\n}$ in \eref{emf}.

By construction ${\cal F}^{\m\n}$  represents a distribution satisfying the Maxwell equations \eref{maxll} and is a Lorentz-covariant tensor. This last property is not at all trivial since our regularization \eref{reg} breaks Lorentz-invariance. There exist actually (under certain aspects) more complicated regularizations than \eref{reg}, which have however the advantage of preserving manifest Lorentz-invariance \cite{LPAM,L1,L2} and which lead to the same result \eref{emf}, see \cite{AL}.

The Maxwell field ${\cal F}^{\m\n}$ is the sum of two non-regular distributions, i.e. distributions that are not represented by functions, the most striking feature being the appearance of the $\delta$-function $P^{\m\n}$.

The field ${\cal F}^{\m\n}$ respects causality. For the contribution ${\cal P}({\cal R}^{\m\n})$ this follows from the fact that it is essentially the pointwise limit of a causal field, i.e. the Lienard-Wiechert field. The term $\tfrac 1 2\,P^{\m\n}$ is supported on the surface $\G^\m$, that gives rise to the new {\it physical} singularity curve $\vec\gamma(b,t)$. From its explicit expression \eref{3sing} one sees that the (measurable) velocity at time $t$ of the point on the curve corresponding to the parameter $b$ -- given by the component of the vector $\partial\vec \gamma(b,t)/\partial t$ orthogonal to the curve $\vec\gamma(b,t)$
-- is $\vec v(t-b)$: the singularity curve propagates therefore with the speed of light, in compatibility with causality. The causality character of the field \eref{emf} is {\it a priori} non trivial, since it could not be derived relying on a causal Green function.

The Hodge-dual of the distributional two-form $P^{\m\n}$ has a simple geometrical interpretation: it is the {\it Poincar\'e-dual} of the surface $\G^\m(b,\la)$, see e.g. \cite{DR}. The relation \eref{dcmn} follows precisely from this property since the boundary of $\G^\m$ is the worldline of the particle.

Equation \eref{dcmn} implies  that in \eref{emf} the electric flux of the massless charge is carried half by the term $\tfrac 1 2\,P^{\m\n}$ and half by the principal part term, since from equations \eref{maxll}, \eref{dcmn} and \eref{emf} it follows that $\pa_\m{\cal P}({\cal R}^{\m\n})=\tfrac12\,j^\n$. Notice, however, that while the field $2{\cal P}({\cal R}^{\m\n})$ satisfies the second Maxwell equation in \eref{maxll}, it would not satisfy the Bianchi identity, because $\partial_{[\m}P_{\n\rho]}\neq0$ too.

For the timelike Lienard-Wiechert field \eref{ftot} the electric flux -- $e$ -- through a sphere centered at the (retarded) particle's position comes entirely from the Coulomb field, while the radiation field can be seen to give a vanishing contribution. On the contrary, once the  distributional limit ${\rm Lim}_{\,V\ra1}$ on $F^{\m\n}$ is taken, the Coulomb field disappears and the electric flux comes entirely from the ``radiation field'' \eref{emf}. This represents a physically as well as mathematically interesting result: when taking the limit from a timelike to a lightlike wordline, the Gauss-law operation and the distributional limit are (maximally) non commuting operations.

As last comment we observe that the energy-momentum tensor $T^{\m\n}$ associated to the Maxwell field \eref{emf}, containing the square of a $\delta$-function is ill-defined, as is the Poynting vector. This, however, does not imply that radiation theory for massless charges is necessarily inconsistent. In fact, even for timelike trajectories the tensor $T^{\m\n}$ -- being the square of the standard Lienard-Wiechert field \eref{ftot} -- is not a distribution and has to be defined through a careful regularization/renormalization process, see \cite{LPAM}. It may be that such a process leads to a well-defined energy-momentum tensor also for lightlike trajectories. Given the explicit expression \eref{emf} it should now be possible to provide for this question a definitive answer.

\vskip0.5truecm

\paragraph{Acknowledgements.}
This work is supported in part by the INFN  Iniziativa Specifica TV12 and by the Padova University Project CPDA119349.

\end{document}